\documentclass[twocolumn,prb,floats,showpacs,superscriptaddress]{revtex4}
\usepackage{graphicx,epsfig}
\usepackage{times}
\usepackage{graphics,dcolumn,bm,float}
\usepackage{amssymb,amsmath,rotate,color}

\begin{document}
\unitlength 1 cm
\newcommand{\be}{\begin{equation}}
\newcommand{\ee}{\end{equation}}
\newcommand{\bearr}{\begin{eqnarray}}
\newcommand{\eearr}{\end{eqnarray}}
\newcommand{\nn}{\nonumber}
\newcommand{\vk}{\vec k}
\newcommand{\vp}{\vec p}
\newcommand{\vq}{\vec q}
\newcommand{\vkp}{\vec {k'}}
\newcommand{\vpp}{\vec {p'}}
\newcommand{\vqp}{\vec {q'}}
\newcommand{\bk}{{\bf k}}
\newcommand{\bp}{{\bf p}}
\newcommand{\bq}{{\bf q}}
\newcommand{\br}{{\bf r}}
\newcommand{\bR}{{\bf R}}
\newcommand{\up}{\uparrow}
\newcommand{\down}{\downarrow}
\newcommand{\fns}{\footnotesize}
\newcommand{\ns}{\normalsize}
\newcommand{\cdag}{c^{\dagger}}

\title{Electric field control of spin-resolved edge states in graphene quantum nanorings}
\author{R. Farghadan}\email{rfarghadan@kashanu.ac.ir}
\affiliation{Department of Physics, University of Kashan, Kashan, Iran }
\author{A. Saffarzadeh}
\affiliation{Department of Physics, Payame Noor University, P.O.
Box 19395-3697 Tehran, Iran} \affiliation{Department of Physics,
Simon Fraser University, Burnaby, British Columbia, Canada V5A
1S6}
\date{\today}

\begin{abstract}
The electric-field effect on the electronic and magnetic
properties of triangular and hexagonal graphene quantum rings with
zigzag edge termination is investigated by means of the
single-band tight-binding Hamiltonian and the mean-field Hubbard
model. It is shown how the electron and spin states in the
nanoring structures can be manipulated by applying an electric
field. We find different spin-depolarization behaviors with
variation of electric field strength due to the dependence of spin
densities on the shapes and edges of this kind of nanorings. In
the case of triangular quantum rings, the magnetization on the
inner and outer edges can be selectively tuned and the spin states
depolarize gradually as the field strength is increased, while in
the case of hexagonal nanorings, the transverse electric field
reduces the magnetic moments on both inner and outer edges
symmetrically and rapidly.
\end{abstract}

\maketitle
\section{Introduction}
The enormous interest in graphene research, both in theory and
experiments is driven by its unusual electronic and magnetic
properties and great potential applications in modern photovoltaic
and nanoelectronic devices
\cite{Novoselov2004,Potasz2009,Novoselov2005,Sanvito2007}. The
fabrication and manipulation of the graphene quantum dots are of
the most desirable developments in this field \cite{Silvestrov}.
Moreover, a carbon-based structure is a good candidate as the
fundamental logic gates for designing ultra-fast spintronic
devices\cite{Wang} and field effect transistors
\cite{Schwierz,Li}.

In graphene quantum dots, magnetic properties may arise from their
topological structure and edge effects. Due to the repulsive
electron-electron interaction in zigzag-edge structures, $\pi$
electrons cannot be paired simultaneously. Therefore, zigzag-edge
graphene nanostructures such as triangular or hexagonal quantum
dots and rings can exhibit magnetic ordering \cite{Farghadan2013,
Fujita} because of an imbalance between the number of atoms
belonging to the two sublattices in graphene bipartite
nanostructures, which produces a net magnetization at the ground
state. In recent years, ring-type nanostructures made of
carbon-based materials like graphene nanoribbons with different
shapes and edges have been studied theoretically \cite{Russo}.
Among them, zigzag-edge graphene quantum rings (GQRs) have a
larger magnetic moment compared to graphene quantum dots due to
more zigzag edges per unit area \cite{Peeters}. In this regard,
magnetic properties, energy levels, and electronic correlations in
graphene rings have also been investigated \cite{Potasz2010,
Bahamon}.

On the other hand, the ability to control electronic and magnetic
properties of material by external electric field is at the heart
of the modern electronic and spintronic devices
\cite{Novoselov2004, Agapito}. In monolayer structures, such as
graphene, the screening of electric field is extremely reduced,
and hence, less energy is consumed. Moreover, graphene quantum
rings and dots with controllable band-gaps, smaller size compared
to the coherence length, and individual electronic and magnetic
properties are more interesting for nanoelectronic applications
\cite{Potasz2011,Saffarzadeh,Palacios}. Therefore, it is essential
to study the effect of electric field in such structures and
several studies by focusing on this effect in graphene quantum
dots with zigzag edges have been reported accordingly
\cite{Zheng2010,Dong,Ma,Zheng2008,Potasz20112}. Moreover,
half-metallicity in graphene nanoribbons induced by electric field
has been predicted \cite{Son}. In these studies, the electronic
and magnetic properties of graphene nanoribbons and quantum dots
by variation of electric field strength have been demonstrated;
however, to our knowledge, the magnetic behavior of GQRs under an
external electric field has not yet been explored.

In this paper, by combining the single-band tight-binding
approximation and the mean-field Hubbard model, we numerically
analyze the influence of transverse electrical-field on electronic
and magnetic properties of narrow triangular and hexagonal GQRs
with zigzag edges. The spin-resolved electronic structure and the
sensitivity of energy gap to the field strength in these nanorings
are investigated. We show that the distribution of local magnetic
moments on the inner and outer edges can be tuned differently by
adjusting the strength of electric field. The evolution of
magnetic moments strongly depends on the type of nanorings, and in
the case of triangular structures, such evolution is even
size-dependent. Therefore, after a short description about our
theory in Section II, we will present our numerical results in
Section III, to see how by applying an electric field rather than
a magnetic field the spin states can be manipulated in such
zigzag-edge nanorings.

\begin{figure}
\centerline{\includegraphics[width=0.9\linewidth]{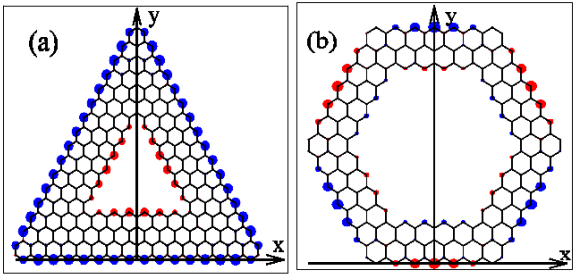}}
\caption{(Color online) Schematic view of the (a) triangular and
(b) hexagonal zigzag-edge graphene rings with distribution of
local magnetic moments in the presence of a uniform electric field
in the $x$-direction. The blue (red) circles correspond to the
spin-up (spin-down) electrons.}
\end{figure}

\section{MODEL AND METHOD}

We simulate the triangular and hexagonal GQRs [shown in Fig.1],
using the $\pi$-orbital tight-binding model and the Hubbard
repulsion treated in the mean-field approximation to include the
effect of electron-electron interaction in the graphene nanoring
calculations, which may induce magnetic localized moments on the
zigzag-shaped edges \cite{Fujita,Farghadan2013}. The Hubbard
Hamiltonian for an $N$-electron ring with $N$ lattice sites in the
presence of a transverse electric field can be written as
\cite{Farghadan2012,Guo}:

\begin{equation}
H=\sum_{i,\sigma}\epsilon_{i,\sigma}n_{i,\sigma}
+t\sum_{<i,j>,\sigma} c^{\dagger}_{i,\sigma}c_{j,\sigma}
+U\sum_{i,\sigma}n_{i,\sigma}\langle n_{i,-\sigma}\rangle \ .
\end{equation}
Here, the operator  $c^{\dagger}_{i\sigma}(c_{i\sigma})$ creates
(annihilates) an electron with spin $\sigma$ at site $i$ and
$n_{i,\sigma}=c^{\dagger}_{i,\sigma}c_{i,\sigma}$ is a number
operator. The first term in Eq. (1) which is defined as
$\epsilon_{i,\sigma}= -ex_{i}E$ describes the effect of electric
field along the $x$-axis, where $e$ is the electron charge, $E$
denotes the electric field strength, and $x_{i}$ is the position
of the $i$th carbon atom along the $x$-direction. The second term
corresponds to the single $\pi$-orbital tight-binding Hamiltonian,
while the third term accounts for the on-site Coulomb interaction
$U$. Accordingly, the magnetic moment (spin) at each atomic site
can be expressed as:
\begin{equation}
m_i= \langle S_i\rangle =(\langle n_{i,\uparrow}\rangle-\langle
n_{i,\downarrow}\rangle)/2 \ .
\end{equation}

Note that, in our calculations the transfer integral between all
nearest-neighbor sites and the Hubbard parameter are set to, $t =
-2.66$ and $ U = 2.82$ eV, respectively.

\section{RESULTS AND DISCUSSION }
In order to study the electronic states and magnetic properties of
the nanorings in the  presence of electric field, we start from an
anti-ferromagnetic spin configuration as an initial condition and
solve the mean-field Hubbard Hamiltonian self-consistently. As
shown in Fig. 1, we consider the electric field distribution along
the $x$-axis for triangular and hexagonal nanorings. The electric
field is generated by two gate electrodes (not shown in the
figure) with opposite electrostatic potentials at the outside of
each ring and parallel to the $y$-axis. These ring structures have
well-defined zigzag shape in both the inner and outer edges. We
describe the number of benzene rings which forms the thickness of
each ribbon by $W$, and the number of carbon atoms in the inner
edge of each ribbon by $L$. The triangular graphene ring with
$W$=3 and $L$=7 consists of 285 carbon atoms. This ring structure
has different number of $A$- and $B$-type atoms on each edge.
Moreover, the inner and the outer edges, each have only one type
of atom. Therefore, the total spin, $S=\sum_im_i$, at zero
electric field, according to the Lieb's theorem \cite{Lieb},
reaches $S=4.5$ with a maximum value of $S = 0.13$. In addition,
the magnetic configurations of the inner and outer edge atoms in
the triangular graphene ring are in the opposite directions with
different distribution of magnetic moment values, as shown in Fig.
1(a). On the other hand, the hexagonal graphene ring consists of
198 carbon atoms, which has the same number of $A$- and $B$-type
atoms. In this structure, the inner and outer edges, each consist
of both types of atoms. Therefore, the total spin value of the
ring is zero and the localized magnetic moments on the both edges
form an antiferromagnetic spin configuration, as shown in Fig.
1(b).
\begin{figure}
\centerline{\includegraphics[width=8.3cm,height=9cm,angle=0]{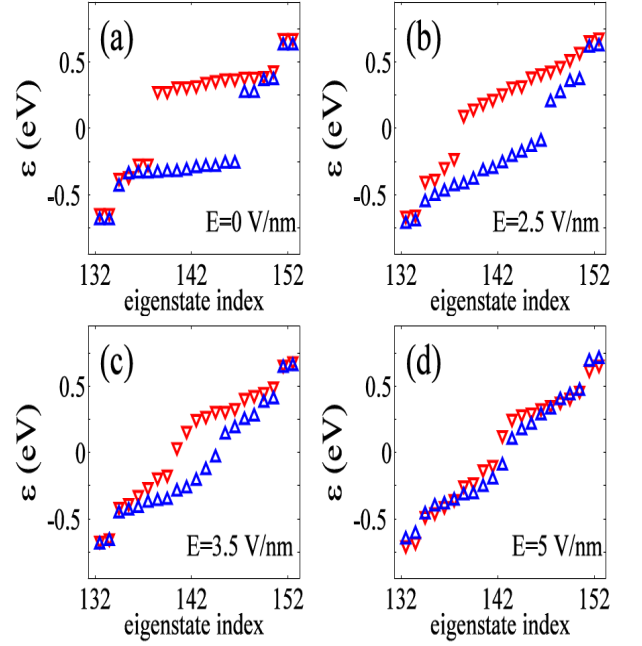}}
\caption{(Color online) Spin-resolved energy levels near the Fermi
energy (set to zero) for the zigzag-edge triangular graphene ring
with $W$=3 and $L$=7 in the presence of different values of
electric field strength. The blue up (down) triangles correspond
to the spin-up (spin-down) electrons.}
\end{figure}

Now we study the single particle energy and the edge magnetic
moment for the case of triangular ring in the presence of electric
field. The strength of the electric field in these calculations is
taken in the range of several volts per nanometer, which is
experimentally achievable if small electrodes are made. In Fig. 2,
the energy spectra for spin-up and spin-down electrons are plotted
near the Fermi energy that is set to zero [Figs. 2(a)- 2(d)].
Without considering the effect of on-site Coulomb repulsion, the
energy spectrum for the edge states at the Fermi energy collapses
to a shell of degenerate states \cite{Potasz2009}. Fig. 2(a) shows
that in the absence of electric field, the electron-electron
interaction opens a gap in the degenerate shell at the Fermi
level. This interaction has a great influence on the degree of
spin polarization of the triangular GQRs and induces a total spin
equal to 4.5. By switching the electric field on, the quasi
degeneracy of edge states is lifted due to the Stark effect and
the energy dispersion increases as shown in Fig. 2(b). In this
electric field strength, the net magnetization remains unchanged.
The spin depolarization, however, occurs when the electrostatic
field is beyond a critical value $E_c$. In the case of triangular
ring (shown in Fig. 1(a)) the critical field is 2.5 V/nm. As the
electric field increases to 3.5 V/nm, the energy levels of spin-up
(spin-down) states, right above (below) the Fermi energy, shift
below (above) the Fermi level and, consequently, the spin gap
between the two spin branches with the same eigenstate index
decreases, and the total spin of the ring reaches $S=2.5$ [Fig.
2(c)]. A further increase in the strength of the electric field
causes a dramatic reduction in the spin gap of the edge states
with the same index which results a remarkable decrease in the
total spin value. As it can be seen in Fig. 2(d), an external
electric field with the strength of 5 V/nm can strongly reduce the
spin gap, and hence, the total spin of the ring reaches 0.5. Our
numerical results show that the total spin of graphene rings is
harder to depolarize than the total spin of graphene quantum dots
\cite{Ma}, indicating more sensitivity to the electric field in
the graphene quantum dots compared to the nanorings.
\begin{figure}
\centerline{\includegraphics[width=1\linewidth]{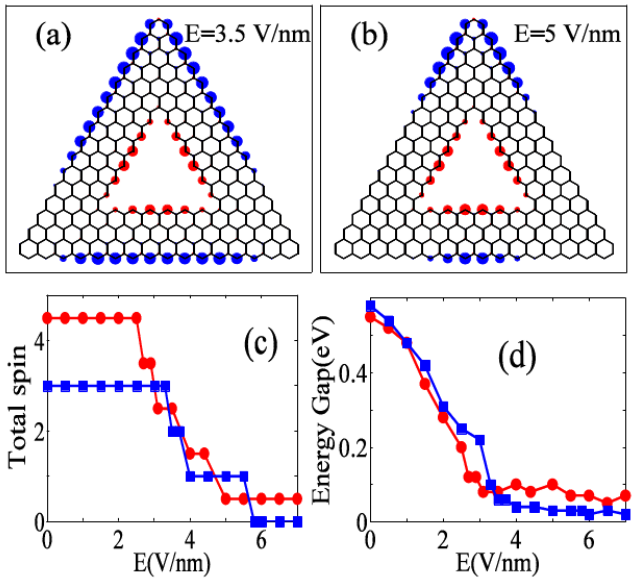}}
\caption{(Color online) Evolution of magnetic edge states with
application of electric fields (a) $E=3.5$ V/nm and (b) $E=5.0$
V/nm in a triangular zigzag GQR. (c) The total spin and (d) the
HOMO-LUMO gaps of two different triangular rings as a function of
electric field strength. The blue squares correspond to a nanoring
with $W=2$ and $L=7$, while the red circles represent a nanoring
with $W=3$ and $L=7$.}
\end{figure}

In order to have a better understanding of the magnetic evolution
of edge states in triangular GQRs as the electric field changes,
we have shown in Fig. 3(a) and 3(b) the distribution of magnetic
moments under the influence of two different values of field
strength. The distribution of magnetic edge states under the
electric field in the range of 0-2.5 V/nm remains unchanged. By
increasing the field strength beyond $E_c=2.5$ V/nm, the magnetic
ordering of edge atoms decreases. The spin-up densities on the
outer edges and far away from the $y$-axis vanish, while the
spin-down densities on the inner edges and the spin-up densities
near the origin on the outer edge of the ring do not change
significantly [Fig. 3(a)]. By further increasing the electric
field strength, the local magnetic moments are restricted to the
center region along the $y$-axis of the ring, as shown in Fig.
3(b).

This magnetic evolution can be understood from Eq. 1.  The
electrostatic potential has low values at atomic sites near the
center line along the $y$-axis in the ring, and the corresponding
on-site energies have small values compared to the on-site energy
of atoms far away from the center region. Therefore, the influence
of electric potential on the atoms located far away from the
center is much stronger than the effect of electron-electron
interaction. However, as the strength of the electric field is
increased above 8 V/nm, the local magnetic moments due to the
on-site Coulomb interaction are dramatically reduced, even for the
inner edge atoms. Accordingly, only the carbon atoms on the outer
edges and near the $y$-axis have nonmagnetic moment with a total
spin equal to 1/2. Note that the net magnetization vanishes only
for the triangular graphene structures with an even number of
atoms. The total spins as a function of electric field strength
for two triangular rings with different sizes are shown in Fig.
3(c). It is clear that the spin depolarization can happen when the
field strength is beyond a critical value. In other words, the
spin depolarization begins to happen when the total spin decreases
by one quantum $\hbar$ as the electric field is increased above a
critical value. This spin reduction manifests itself as a steplike
function of the electric field strength. The electric field cannot
completely depolarize the total spin of a triangular ring with
$W=3$ and $L=7$, and such a structure has the minimal value
$S=1/2$ due to the odd number of electrons. On the other hand, the
complete spin depolarization occurs for a triangular carbon ring
with $W=2$ and $L=7$. This triangular structure consists of 186
carbon atoms which indicates an even number of electrons at half
filling, and hence, the total spin completely depolarizes, i.e.,
$S=0$. In our calculations, the critical value of the electric
field for starting the spin depolarization in the larger (smaller)
triangular ring is $E=2.5$ V/nm (3.5 V/nm) and the minimum spin
can reach $E=5$ V/nm (6 V/nm). This means that the larger GQR,
which is affected by a larger electrostatic potential, depolarizes
easier than the smaller ring. Accordingly, the spin depolarization
of such triangular rings depends on the number of electrons at
half filling and also the field strength. To investigate the field
effect on the electronic structure of the nanorings, we have shown
in Fig. 3(d) the energy gap as a function of field strength for
the two different triangular rings. It is clear that the highest
occupied molecular orbital (HOMO) and the lowest unoccupied
molecular orbital (LUMO) gap reduces as the field strength
increases. For instance, in the case of thinner ring, the energy
gap decreases to the small value of ~20 meV at $E\geq$ 5 V/nm.
However, in contrast with the rectangular graphene nanodots,
\cite{Zheng2008} in which the nanodots switch from
antiferromagnetic to diamagnetic ground state at low electric
fields, and the HOMO-LUMO gap becomes spin-dependent, the ground
state of the triangular nanorings remains antiferromagnetic even
at high electric fields such as $E=5$ V/nm. This indicates the
strong influence of the geometry, quantum size, and edges on the
induced magnetic moments in this type of nanoring compared to
those in the square nanodots \cite{Zheng2008}.

\begin{figure}
\centerline{\includegraphics[width=8.1cm,height=4.2cm,angle=0]{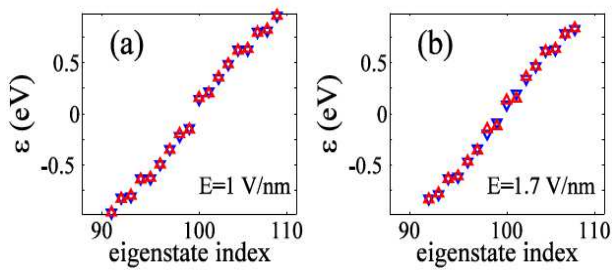}}
\caption{(Color online) Spin-resolved energy levels near the Fermi
energy (set to zero) for zigzag-edge hexagonal GQRs in the
presence of electric field (a) $E=1$ V/nm, and (b) $E=1.7$ V/nm.
The blue up-triangles and red down-triangles correspond to the
spin-up and spin-down electrons, respectively.}
\end{figure}

Now we consider the field effect on the electronic and magnetic
properties of the hexagonal nanoring with zigzag edge termination,
shown in Fig. 1(b). In Fig. 4, we have depicted the spin-resolved
energy spectra for spin-up and spin-down electrons as the strength
of electric field is increased. This kind of nanorings does not
show any spin gaps in its electronic spectra and the net
magnetization remains zero in both the presence and absence of the
applied field. On the contrary, the local magnetic moment on each
side of the hexagonal ring is nonzero, even under the influence of
electric field. To see this behavior, we have shown in Fig. 5(a)
and 5(b) the distribution of induced magnetic moments in the
presence of two field strengths $E=1.7$ V/nm and $E=1.8$ V/nm,
respectively. It is clear that, at $E=1.7$ V/nm, the spin states
show the same evolution on each side (edge) of the six inner or
outer edges of the hexagonal ring. Moreover, the local magnetic
moments on each side of the ring vanish rapidly when the electric
field strength varies from 1.7 to 1.8 V/nm. Comparing the Figs.
5(a) and 5(b), one can conclude that the electronic structure of
the hexagonal ring is more sensitive to the field strength,
compared to the triangular structures, and its spin depolarization
occurs at lower values of electric field (see Figs. 3(a) and 3(b)
and 5(a) and 5(b) for comparison). We find that, applying a
transverse electric field, even on a small part of the hexagonal
ring, reduces the spin states symmetrically on the zigzag edge
atoms [not shown here].

\begin{figure}
\centerline{\includegraphics[width=1\linewidth]{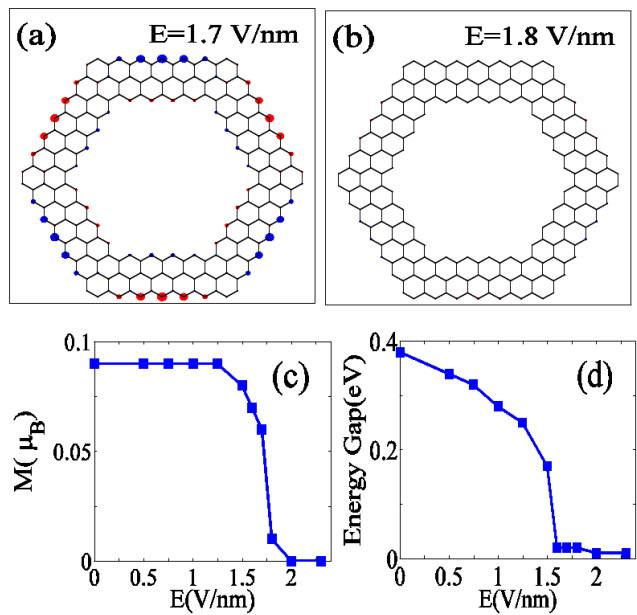}}
\caption{(Color online) Evolution of magnetic edge states in a
hexagonal GQR in the presence of electric fields (a) $E=1.7$ V/nm
and (b) $E=1.8$ V/nm . (c) The largest magnetic moment ($M$) of
edge atoms and (b) the HOMO-LUMO gap of the hexagonal GQR as a
function of the electric field strength.}
\end{figure}

In Fig. 5(c), the maximum value of the local magnetic moments
($M=\mu_BS$) in the hexagonal ring is shown as the electric field
is varied. It can be seen that the magnetic moment is relatively
small, compared to the triangular rings. Since the value of $S$ is
less than 1, the spin depolarization does not behave like what was
observed in the triangular rings, in which the depolarization
occurs when the electric field strength is beyond some critical
values. Consequently, the local magnetization in the hexagonal
systems is easier to depolarize than the local magnetic moment in
the triangular structures. We have also shown the behavior of the
energy gap as a function of field strength in Fig. 5(d). Our
results reveal that, although the HOMO-LUMO gap of the hexagonal
rings responds to the variation of field strength very similar to
that of the triangular rings, it drops and vanishes much faster
than those in Fig. 3(d).

\section{Conclusion}
In summary, the influence of electric field and electron-electron
interaction on the edge states in the zigzag-edge GQRs was studied
by means of the single-band tight-binding Hamiltonian and the
mean-field Hubbard model. There is a competition between the
electron-electron interaction and the external electric field to
depolarize the local magnetic moments on the inner and outer edges
of the nanorings. We showed that the strength of electrostatic
potential strongly affects the magnetization in our systems. The
presence of electrostatic field mainly affects the electronic
spectrum in the vicinity of the Fermi energy, while the single
particle states which are located far away from this energy,
remain unchanged and therefore the magnetic properties of
nanorings are determined by the $\pi$-electrons around the Fermi
level.

In the case of triangular rings, the inner and the outer edge
states show different evolutions under the external electric
field, and the net spin-polarization of the nanorings might be
enhanced by choosing appropriate field strengths such as the
non-uniform electrostatic fields proposed in ref\cite{Dong}. On
the contrary, the evolution of the magnetic moments on all of the
edges (whether they be the inner edges or the outer edges) of the
hexagonal nanorings is the same by variation of field strength,
which clearly indicates that the size and the shape of the
nanorings are crucial factors in the electric-field-induced
depolarization in this type of nanorings. Although by increasing
the size of the nanorings many particle interactions are expected
to be more important \cite{Potasz2010}, these findings can still
be useful to qualitatively understand the variation of energy
spectra and spin depolarization under the influence of electric
field in larger graphene nanorings.

\section*{Acknowledgement}
This work financially supported by university of Kashan under
grant No. 228762.

\end{document}